%% file: 12_02_01_manuscript.tex
\documentclass[number,12pt]{elsarticle}
\usepackage{multirow}
\usepackage{epsfig, setspace, pmgraph}%doublespace}
\usepackage[lofdepth,lotdepth]{subfig}
\usepackage{graphicx}
\usepackage{afterpage}
\usepackage{amsmath, amsfonts, amssymb,commath}
\usepackage{dsfont,courier}
\usepackage{verbatim}
\usepackage{upgreek}
\usepackage{longtable}
\usepackage{hyperref}
\usepackage{color,soul}
\usepackage{lineno}
\usepackage{caption}
\usepackage{booktabs}

\journal{Materials Science \& Engineering A}

\begin{document}

\begin{frontmatter}

\title{Constitutive behavior of as-cast A356}

\author[ubc]{M. J. Roy\corref{cor1}}
\ead{majroy@interchange.ubc.ca}
\author[ubc]{D. M. Maijer}
\ead{daan.maijer@ubc.ca}
\author[ENSMA]{L. Dancoine}
\ead{louise.dancoine@etu.ensma.fr}

\address[ubc]{Dept. of Materials Engineering, The University of British Columbia, Vancouver, BC, Canada V6T 1Z4}
\address[ENSMA]{Institut PPRIME - CNRS - Universit\'{e} de Poitiers - ENSMA - UPR 3346 - D\'{e}partment M\'{e}canique des Mat\'{e}riaux - T\'{e}l\'{e}port 2 - 1 Avenue Cl\'{e}ment Ader - BP 4019 - 86961 FUTUROSCOPE CHASSENEUIL Cedex - France}

\cortext[cor1]{Corresponding author, Tel. +1 604 822 2676;
Fax +1 604 822 3619}

\begin{abstract}
The constitutive behavior of aluminum alloy A356 in the as-cast condition has been characterized using compression tests performed over a wide range of deformation temperatures (30-500$^{\circ}$C) and strain rates ($\sim$0.1-10 s$^{-1}$). This work is intended to support the development of process models for a wide range of conditions including those relevant to casting, forging and machining.  The flow stress behavior as a function of temperature and strain rate has been fit to a modified Johnson-Cook and extended Ludwik-Hollomon expression. The data has also been assessed with both the strain-independent Kocks-Mecking and Zener-Hollomon frameworks. The predicted plastic flow stress for each expression are compared. The results indicate that the extended Ludwik-Hollomon is best suited to describe small strain conditions (stage III hardening), while the Kocks-Mecking is best employed for large strain (stage IV). At elevated temperatures, it was found that the Zener-Hollomon model provides the best prediction of flow stress.
\end{abstract}


\end{frontmatter}
%\begin{linenumbers}
%\input{11_10_20_Highlights.tex}
\input{12_02_03_Nomenclature.tex}
\section{Introduction}\label{sec:intro}
Cast aluminum alloys continue to displace ferrous alloys in transport applications where weight and part count reduction are important considerations. The manufacturing processes for commonly cast aluminum alloy A356 (Al�7Si�0.3Mg) components often includes machining operations prior to final heat treatment.  Modelling and optimizing processes occurring before heat treatment requires a comprehensive characterization of the flow stress under a wide range of thermal and deformation conditions. Processing that occurs in this time frame affects the final component properties. This includes casting which involves low strains and strain rates at high temperatures, to forging which involves high strains and strain rates at intermediate temperatures.

There have been many phenomenological and physically-based constitutive models that have been employed to predict plastic flow stress as a function of material structure, temperature and strain rate. While a significant number of studies have been undertaken to study the deformation behavior of A356, all have focused on the material following various post-casting thermal treatments. The most common heat treatment for A356 is the T6 heat treatment, which involves bringing the secondary phases into solution, quenching and then artificially ageing. As such, the deformation behaviour of A356 at room temperature has been well characterized in the T6 condition \cite{Wang-T6.04,Ran-T6.06,Lee.07,Zhang.96}, and at elevated temperatures in the solution-treated state \cite{Estey.04}. The response of A356 during semi-solid forming \cite{Ashouri.08} and rheocasting \cite{Lashkari.08} have also been studied but the non-typical microstructure produced during these processes make this data of limited use for traditionally cast A356.

In order to address the lack of deformation data for A356 at elevated temperatures and strain rates in the as-cast condition, a large number of isothermal compression tests were conducted in an effort to develop a comprehensive constitutive equation.  Prior to presenting the experimental results, a literature review of suitable constitutive expressions will be presented. Analysis of the experimental data within each constitutive framework will follow with discussion of efficacy and performance. This will be accomplished by comparing the experimental flow stresses to those predicted with the goal of identifying the best approach for a given thermomechanical state.
\section{Constitutive expression overview}\label{sec:ConstitutiveExpr}
In general, the phenomenological deformation behavior of aluminum alloys is dominated by strain hardening at low temperatures, which transitions to a time or strain-rate dependent (creep) response at elevated temperatures.  Models of aluminum casting processes consider the material either from an elastic time-dependent creep standpoint, or as an elastic-plastic strain-dependent material. The former formulations considered the Sellars-Tagart \cite{Sellars.66} or Garafalo relationships \cite{Garafalo.63,AlhassanAbu.03,Drezet.96}, while the latter \cite{VanHaaften.99,Phillion.08,Drezet.10} used extended Ludwik-Hollomon expressions. Constitutive behaviors of as-cast wrought aluminum AA1050, AA3104 and AA5182 have also been described with a simultaneous combination of time and strain dependence \cite{vanHaaften.02}. The Johnson-Cook \cite{Johnson-Cook.83} model, which is specifically structured to isolate strain-rate and temperature dependence, is another popular phenomenological approach although it has not been employed for cast materials. Since there is often a co-dependance of temperature and strain-rate, many have modified the original Johnson-Cook model to better describe specific materials at specific temperatures and strain-rate ranges \cite{Samantaray.09,Maheshwari.10,Lin.10}. While being quite effective in predicting flow stresses, the main drawback of using phenomenological approaches is that they are only valid within the experimental strain range from which the constitutive expression has been derived.

Physically-based models, such as the Zerilli-Armstrong \cite{Zirelli-Armstrong.87} which considers grain size or the more general Kocks-Mecking \cite{Mecking.81,Estrin.84,Follansbee-MTS.88,Kocks.03}, employ material parameters to predict flow stress. Grain size in cast Al-Si alloys has been shown to have little effect on strength compared to other microstructural features such as dendrite spacing \cite{Rooy.11,Maijer.04,Goulart.06}. As a result, models which use grain size as their underlying basis are not well suited to material with dendritic microstructure. Based on its more fundamental derivation, the Kocks-Mecking approach has been selected as a candidate physically-based model for as-cast A356 in the current work.

\subsection {Kocks-Mecking}\label{sec:KM_overview}
The Kocks-Mecking hardening model is the most common physically-based approach to describe constitutive behavior. The underlying premise of the Kocks-Mecking hardening model is that the work hardening rate $\Theta=\dif \sigma/\dif \varepsilon$ approaches zero at some saturation stress, $\sigma_s$ which is a function of strain rate and temperature. As a result, plotting $\Theta/\mu$ versus $\sigma/\sigma_s$ creates a master curve that is accurate over a large range of hardening such that:
\begin{equation}\label{eq:KMbaseline}
\frac{\Theta}{\Theta_0}=f\left(\frac{\sigma}{\sigma_s}\right)
\end{equation}
Typically, Eq. \ref{eq:KMbaseline} is assumed to have the form of a linear Voce-type relationship:
\begin{equation}\label{eq:KMlinear}
\Theta=\Theta_0\left(1-\frac{\sigma}{\sigma_s}\right)
\end{equation}
with an integrated form of:
\begin{equation}\label{eq:KMbaselineflow}
\frac{\sigma-\sigma_s}{\sigma_y - \sigma_s}=\exp\left(\frac{-\Theta_0\varepsilon}{\sigma_s}\right)
\end{equation}
The saturation stress, $\sigma_s$, is expressed as:
\begin{equation}\label{eq:KM_sigS}
\sigma_s=\sigma_{s0}\frac{\mu}{\mu_0}\left(1-\left(\frac{g}{g_0}\right)^{1/q}\right)^{1/p}
\end{equation}
which is a function of a normalized activation energy term $g$ is defined as \cite{Kocks.03}:
\begin{equation}\label{eq:KM_g}
g=\frac{kT}{\mu \bar{b}^3}\ln{\frac{\dot \varepsilon_0}{\dot \varepsilon}}
\end{equation}
where $\dot \varepsilon_0$ is the minimum strain rate that converges all data to a single function that relates $\sigma_s/\mu$ to $g$. The physical basis of $p$ and $q$ ($0\leq p\leq 1$ and $1\leq q\leq 2$) in Eq. \ref{eq:KM_sigS} represent the shape of dislocation obstacle profiles \cite{Follansbee-MTS.88}, which in turn decide values of $\sigma_{s0}$ and $g_0$. These values are phenomenological in nature \cite{Kocks.03,Kocks.75}, and have not been used to distinguish between discrete obstacle types such as precipitates or grain boundaries. The yield strength, $\sigma_y$, can also be expressed as a function of $g$, which renders Eq. \ref{eq:KMbaselineflow} a $\sigma$-$\varepsilon$ relationship that is a function of strain rate and temperature.
\subsection {Johnson-Cook}\label{sec:JC_overview}
The Johnson-Cook model remains a widely adopted phenomenological model, potentially owing to simplicity and not requiring an extensive experimental regimen to develop. The heuristic basis of the Johnson-Cook model is such that the flow stress at elevated temperatures is determined by scaling the constitutive behavior at low temperatures and quasi-static strain rates. This approach consists of determining the product of three terms which is typically expressed as:
\begin{equation}
\begin{split}
\hat \sigma_{\text{JC}}&=\sigma^{\ast}(\epsilon)f(\dot \varepsilon)f(T)\\
            &=\sigma^{\ast}(\epsilon)\left(1+C\ln\left(\frac{\dot \varepsilon}{\dot \varepsilon^{\ast}}\right)\right)\left(1-T_{\text{H}}^m\right)\label{eq:stdJC}
\end{split}
\end{equation}
The reference stress $\sigma^{\ast}(\epsilon)$ in Eq. \ref{eq:stdJC} is a relationship that describes the quasi-static flow stress at a low temperature as a function of strain, usually following a standard Ludwik-Hollomon expression, $\sigma^{\ast}(\epsilon)=\sigma_{y}^{\ast}+K^{\ast}\epsilon^{n^{\ast}}$. The reference stress data is collected at reference strain rate $\dot \varepsilon^{\ast}$ (usually 1 s$^{-1}$) at temperature $T^{\ast}$. The second term describes the strain rate scaling based on a fitting coefficient $C$, and the third term is intended to capture the effects of temperature as a function of the homologous temperature, $T_{\text{H}}$ and a constant $m$ ($\simeq 1$ for aluminum \cite{Johnson-Cook.83}) where:
\begin{equation}\label{eq:homoTemp}
T_{\text{H}}=\frac{T-T^{\ast}}{T_{\text{melt}}-T^{\ast}}
\end{equation}
The main drawback of the standard Johnson-Cook model is that strain hardening is not explicitly affected by strain rate or temperature. The standard Johnson-Cook model has been modified by various functions of strain rate and temperature by other researchers to accommodate for this \cite{Liang.99}. In this study, the Johnson-Cook framework has been utilized to describe the behavior of A356.
\subsection{Extended Ludwik-Hollomon}\label{sec:LH_overview}
Another phenomenological model frequently employed for cast aluminum alloys is the extended Ludwik-Hollomon expression. In order to capture the simultaneous evolution of both strain hardening and strain rate effects with temperature for aluminum, van Haaften \cite{vanHaaften.02} fit stress-strain data to an extended Ludwik-Hollomon expression of the form:
\begin{equation}
\hat \sigma_{\text{LH}}=K\left(\varepsilon^0+\varepsilon\right)^N\left(\dot\varepsilon^0+\dot\varepsilon\right)^M
\end{equation}
where $K$, $N$ and $M$ are constants corresponding to strength, strain hardening and strain-rate sensitivity, respectively. The constants $\varepsilon^0$ and $\dot \varepsilon^0$ were reported by van Haaften to be numerically necessary in order to provide the correct yield strength at room temperature using the Simplex method for fitting. Expanding on the work of Phillion et al. \cite{Phillion.08,Drezet.10}, this expression has been modified such that $K$, $M$ and $N$ are functions of temperature:
\begin{equation}\label{eq:LHbaseline}
\hat \sigma_{\text{LH}}=K(T)\varepsilon^{N(T)}\left(\frac{\dot \varepsilon}{\dot \varepsilon_1}\right)^{M(T)}
\end{equation}
where $\dot \varepsilon_1$ is a normalization strain rate of 1 s$^{-1}$ . The advantage of this phenomenological approach as opposed to the Johnson-Cook model is that $N(T)$ and $M(T)$ are both independent functions which when factored from the flow stress enable $K(T)$ to be determined via regression. Compared to the single independent function that forms the basis of the Johnson Cook model, this is a more direct approach.
\subsection{Zener-Hollomon}\label{sec:ZH_overview}
The Zener-Hollomon is an approach that has been widely used to characterize strain-independent, high-temperature flow stress. The flow stress is expressed as a function of the Zener-Hollomon parameter $Z$:
\begin{equation}\label{eq:ZHsig}
\hat \sigma_{\text{ZH}}=\frac{1}{\rho}\ln\left(\left(\frac{Z}{\Phi}\right)^{1/\eta}+\sqrt{\left(\frac{Z}{\Phi}\right)^{2/\eta}+1}\right)
\end{equation}
where $Z$ is expressed in terms of strain rate and temperature \cite{Zener.44}:
\begin{equation}
\label{eq:ZenerEps}Z=f(\dot \varepsilon, T)=\dot \varepsilon \exp\left(\frac{Q}{RT}\right)
\end{equation}
or stress by using the Sellars-Tagart\cite{Sellars.66} Arrhenius-type equation:
\begin{equation}
Z=\Phi \sinh \left(\rho \sigma\right)^\eta\label{eq:ZenerSig}
\end{equation}
where $Q$ is an activation energy, $R$ is the gas constant, $\sigma$ is the strain-independent flow stress and fitting coefficients are $\Phi$, $\rho$ and $\eta$.  These latter three coefficients are occasionally reported to be solute-dependent \cite{vanHaaften.02}, however studies on steel have suggested that they are merely fitting parameters and have no physical basis \cite{Medina.96}. The only parameter with a physical basis is the activation energy $Q$, which is an activation enthalpy.

Owing to the wide range of constitutive expressions available, the preceding expressions will be applied to the experimental results described in the next section to determine the best assessment of the constitutive behavior of as-cast A356.
\section{Material and experimental methodology}\label{sec:Material}
Al-Si-Mg alloys, such as A356, have an as-cast microstructure consisting of primary aluminum dendrites ($\alpha$-Al) which form during the initial solidification phase, surrounded by an Al-Si eutectic. As-cast microstructure refinement, and a corresponding strength increase, is achieved primarily through decreasing solidification time or which decreases the secondary dendrite arm spacing (SDAS). Precipitation strengthening particles (Mg$_2$Si) occur throughout the microstructure, and other tertiary phases may be present due to impurities (Fe,C). The T6 heat treatment typically applied to this alloy serves to modify the eutectic structure and refine/redistribute the Mg$_2$Si particles.

\begin{figure}
\centering
      \includegraphics[width=\textwidth]{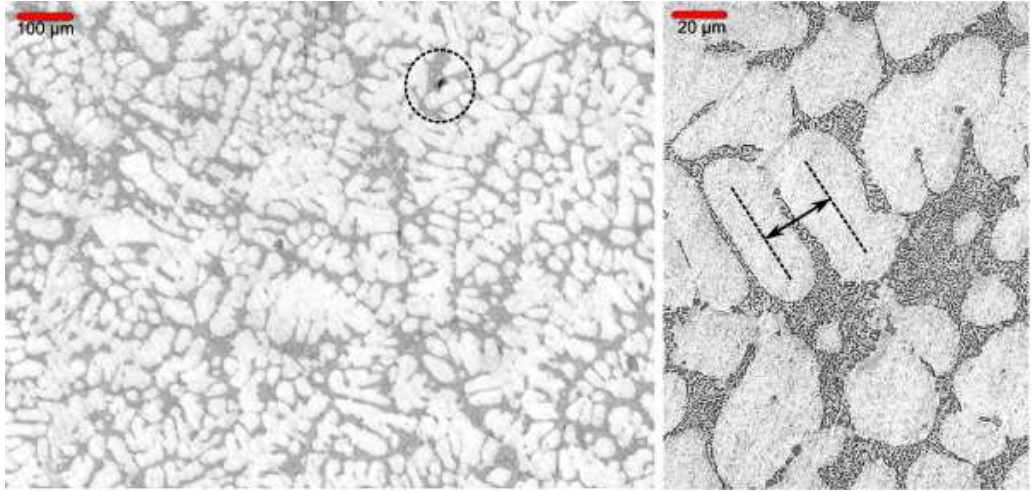}
      \caption[]{As-cast A356 microstructure showing $\alpha$-Al (bright), Al-Si eutectic (dark) phases. An example of a typical pore has been highlighted on the left and the high resolution of the microstructure on the right has been annotated to show the secondary spacing (SDAS).}
      \label{fig1}
\end{figure}

The material investigated in this study is a strontium-modified A356 cast from metal extracted from the melt supply of a North American aluminum alloy wheel manufacturer with the nominal chemical composition given in Table \ref{table:composition}. The melting temperature ($T_{\text{melt}}$) of this alloy is $612.5^{\circ}$C \cite{Rooy.11}. Samples of this material were cast in a permanent steel mold with a water-cooled chill. Figure \ref{fig1} shows an optical micrograph of the typical as-cast microstructure.

\begin{table}[h!]
\centering
\caption{A356 composition in wt-\%\label{table:composition}}
    \begin{singlespacing}
    \begin{tabular}{lllll}
    \hline
    Element & Si & Mg & Na & Sr\\
    Range (wt-\%) & 6.5-7.5 & 0.25-0.4 & $\sim$0.002 & $\sim$0.005\\
    \hline
\end{tabular}
\end{singlespacing}
\end{table}

While the cooling rate was not measured during solidification, the SDAS and porosity were quantified with samples extracted at regular intervals from the casting. Quantification was conducted via optical microscopy and Clemex Vision PE software. Measurements were made at 12 locations representing a composite area greater than 85 mm$^2$. The SDAS, defined as the average distance between the approximated centers of active secondary dendrite arms \cite{Grugel.93}, was manually measured on both equiaxed and columnar dendrites.  The average SDAS was found to be 39.9 $\upmu$m with a standard deviation of 7.0 $\upmu$m. Both the average and standard deviation are based on 491 discrete arm measurements (Fig. \ref{fig1}) distributed over the 12 locations; each location comprised at least 30 measurements. The measured gas-based porosity was widely dispersed with a mean percent area of 0.154\%. This SDAS and porosity are representative of those observed in industrial castings from this manufacturer.

\begin{figure}[h!]
\centering
      \includegraphics[width=2 in]{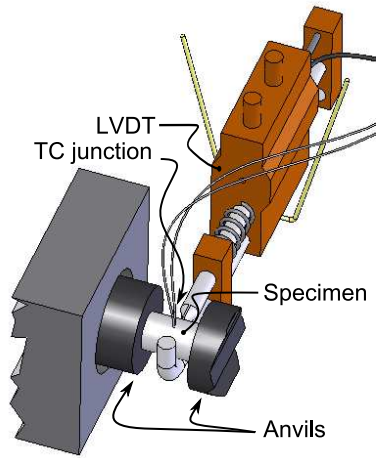}
      \caption[]{A cutaway depiction of the Gleeble experimental arrangement. The specimen with a thermocouple junction is positioned between the anvils and LVDT.}
      \label{fig2}
\end{figure}

Cylindrical compression test specimens, nominally measuring 10 mm in diameter by 15 mm in length, were extracted from the as-cast sample. Deformation tests were then conducted on a Gleeble\footnote{Gleeble is a trademark of Dynamic Systems, Inc., Poestenkill, NY.} 3500 thermomechanical simulator fitted with isothermal tungsten carbide anvils. A small amount of nickel-based lubricant was placed between the sample and the anvils. During each test, the temperature was controlled with a type-K thermocouple mounted to the center of each specimen. Instantaneous diametral deformation was measured with a linear voltage displacement transducer (LVDT) actuated by quartz rods. This experimental arrangement is shown in Fig. \ref{fig2}. The procedure followed for each test was:
\begin{enumerate}
    \item The sample was first loaded between the platens with a pre-load of less than 0.5 kN.
    \item The hydraulic actuator was retracted by 2 mm to allow for thermal expansion. Pre-load on the sample was maintained via friction.
    \item The sample was Joule heated at a rate of 5$^{\circ}$C s$^{-1}$ and then held at the target test temperature for 60 seconds.
    \item The sample was deformed at the prescribed deformation rate to target strains of 0.2, 0.4 and 0.55 based on the specimen's nominal length at ambient temperature.
    \item Based on the prescribed target strain rate, the data acquisition rate from the load cell, thermocouple and LVDT was greater than 200 samples per unit strain during deformation.
    \end{enumerate}
    
At elevated strain rates, heat generated during the deformation was greater than could be removed by the anvils. This caused a slight increase ($\sim10^{\circ}$C) in the average deformation temperature $T_a$, and was more predominant at lower temperatures and higher strain rates. Although the pre-load at all temperatures was kept below 0.5 kN, creep occurred at elevated temperatures during the hold stage prior to deformation. While the creep rate differs with temperature, the amount of creep was nearly identical from test to test as the pre-deformation heating cycle was the same. Differing amounts of thermal expansion caused by variations in specimen geometry render quantification of the amount of creep prior to deformation intractable. However, due to the low load and brief thermal history each sample experienced pre-deformation, this amount of creep is small and has been considered part of the thermal expansion included in the initial diameter, $D_0$.

Uniform plastic deformation was assumed in all cases such that the instantaneous true strain $\varepsilon$ was found from the instantaneous diameter $D$ according to:
    \begin{equation*}
    \varepsilon=-2\ln\frac{D}{D_0}
    \end{equation*}
The instantaneous strain record was then differentiated to provide the instantaneous strain rate, $\dot \varepsilon$. The average strain rate, $\dot \varepsilon_a$, calculated as the mean strain rate during deformation, was used to categorize each test. This is also true for the average temperature, $T_a$, as well. All experimental data points of temperature, load, strain and strain rate have been incorporated in the following analysis.
\section{Experimental Results}\label{Sec:ExpResults}
\begin{figure}[h!]
\centering
      \includegraphics[width=3.75 in]{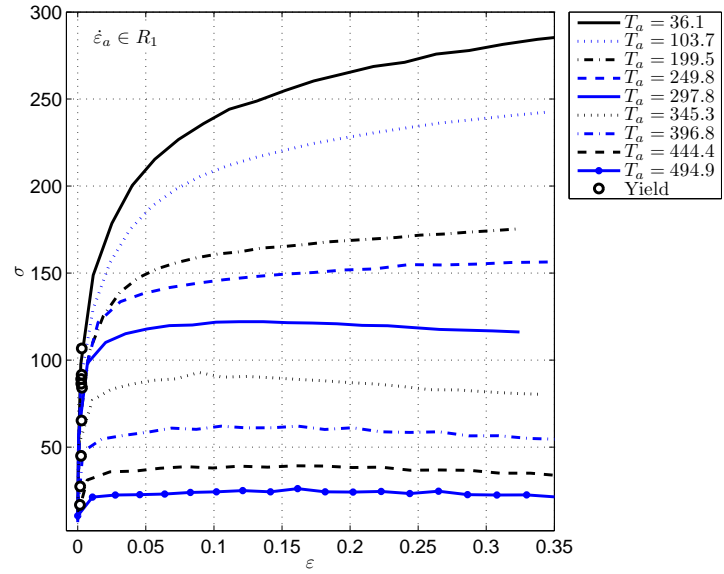}
      \caption[]{Characteristic measured stress-strain results for $\dot \varepsilon \in R_1=$ 0.06 to 0.11 s$^{-1}$.\label{fig:lowspeed}}
      \label{fig3}
\end{figure}

\begin{figure}[h!]
\centering
      \includegraphics[width=3.75 in]{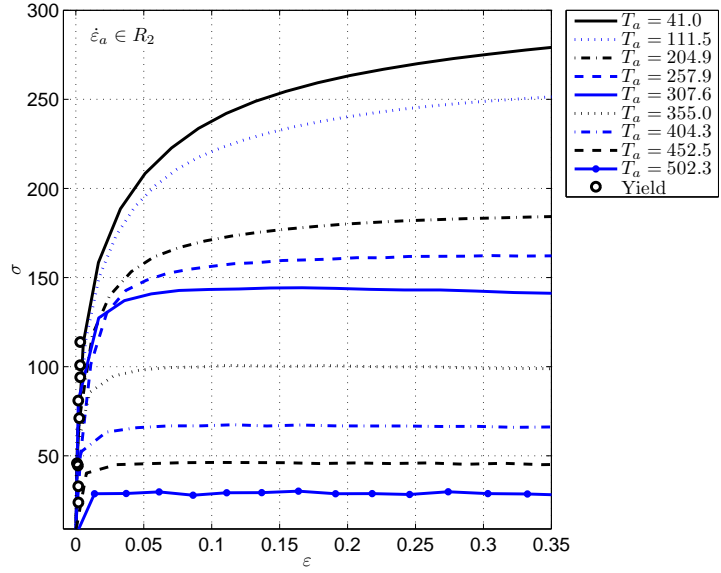}
      \caption[]{Characteristic measured stress-strain results for $\dot \varepsilon \in R_2=$ 0.57 to 0.98 s$^{-1}$.\label{fig:midlowspeed}}
      \label{fig4}
\end{figure}

\begin{figure}[h!]
\centering
      \includegraphics[width=3.75 in]{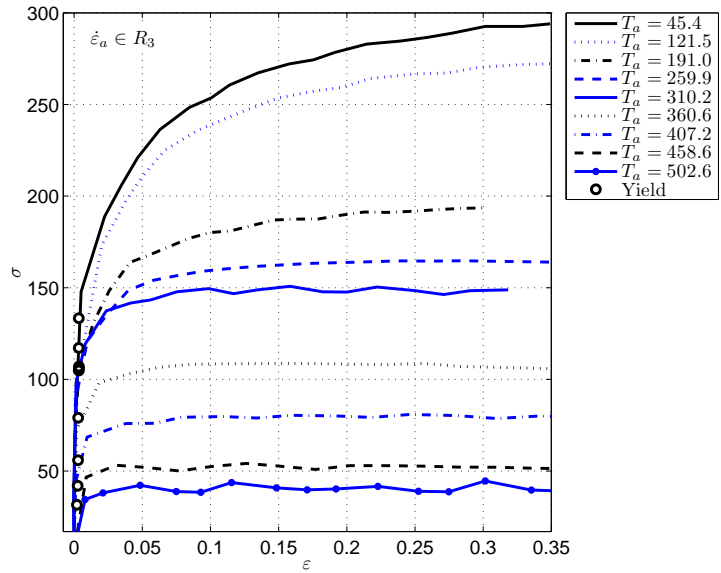}
      \caption[]{Characteristic measured stress-strain results for $\dot \varepsilon \in R_3=$ 3.59 to 5.33 s$^{-1}$.\label{fig:midhispeed}}
      \label{fig5}
\end{figure}

\begin{figure}[h!]
\centering
      \includegraphics[width=3.75 in]{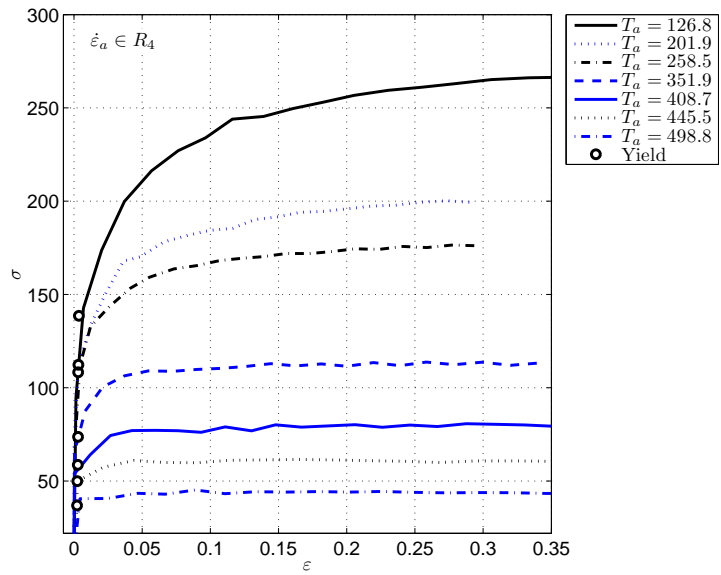}
      \caption[]{Characteristic measured stress-strain results for $\dot \varepsilon \in R_4=$ 7.58 to 12.24 s$^{-1}$.\label{fig:hispeed}}
      \label{fig6}
\end{figure}

In total, 55 successful compression tests were carried out at a range of temperatures, strains and strain rates. Target strains and strain rates were calculated based on the nominal length of the specimen at ambient temperature. Temperatures of 30$^{\circ}$C, 100$^{\circ}$C, and between 200 to 500$^{\circ}$C in increments of 50$^{\circ}$C and strain rates of 0.1, 1, 5 and 10 s$^{-1}$ were targeted. Following the tests, the strain rates achieved were calculated based on the strain versus time information. The achieved strain rates based on each target strain rate were grouped in ranges of 0.06 to 0.11 s$^{-1}$, 0.57 to 0.98 s$^{-1}$, 3.59 to 5.33 s$^{-1}$ and 7.58 to 12.24 s$^{-1}$, which are referred to further as $R_1$, $R_2$, $R_3$ and $R_4$, respectively. Flow curves are presented for each of these ranges of $\dot \varepsilon_a$ in Figures \ref{fig3}-\ref{fig6}, while all test conditions are summarized in Table \ref{table:Results1}. The variation in target strain rates and strains was attributed to thermal expansion and variation in specimen length. A number of tests at the highest target strain rates in each temperature range were not successful owing to non-uniform deformation and barreling. Tests where these conditions developed were defined as failed tests and their data is not shown nor used for subsequent analysis.

\addtolength{\belowcaptionskip}{0pt}
\begin{table}[!h]
 \caption{$T_a$, $\dot \varepsilon_a$, $\sigma_y$ and $\varepsilon_p$ attained for all tests}
\centering
\small
\captionsetup[subfigure]{font=small,skip=-1pt}
\vspace{-0.125 in}
\subfloat[$\sim30^{\circ}$C]{
\begin{tabular}{ccccc}
\hline No. & $T_a$& $\dot \varepsilon_a$   & $\sigma_y$& $\varepsilon_{p}$\\
\hline
1 & 33.4 & 0.06 &  97 & 0.13\\
2 & 36.1 & 0.07 & 112 & 0.37\\
3 & 34.0 & 0.07 & 109 & 0.18\\
4 & 41.0 & 0.57 & 114 & 0.42\\
5 & 38.1 & 0.71 & 102 & 0.19\\
6 & 36.8 & 0.71 & 105 & 0.19\\
7 & 37.5 & 0.72 & 113 & 0.19\\
8 & 33.7 & 0.74 & 107 & 0.14\\
9 & 45.4 & 3.95 & 137 & 0.35\\
\hline
\end{tabular}}
\subfloat[$\sim100^{\circ}$C]{
\begin{tabular}{ccccc}
\hline No. & $T_a$& $\dot \varepsilon_a$   & $\sigma_y$& $\varepsilon_{p}$\\
\hline
10 & 103.7 & 0.07 &  95 & 0.36\\
11 & 102.4 & 0.07 & 104 & 0.17\\
12 & 104.3 & 0.70 & 120 & 0.14\\
13 & 112.4 & 0.74 & 102 & 0.45\\
14 & 106.9 & 0.77 & 106 & 0.20\\
15 & 104.7 & 0.77 & 117 & 0.15\\
16 & 106.6 & 0.78 & 111 & 0.19\\
17 & 121.5 & 4.23 & 121 & 0.45\\
18 & 126.9 & 8.27 & 117 & 0.39\\
\hline
\end{tabular}}\\
\subfloat[$\sim200^{\circ}$C]{
\begin{tabular}{ccccc}
\hline No. & $T_a$& $\dot \varepsilon_a$   & $\sigma_y$& $\varepsilon_{p}$\\
\hline
19 & 201.3 & 0.08 &  85 & 0.19\\
20 & 199.5 & 0.08 &  84 & 0.32\\
21 & 204.9 & 0.72 &  84 & 0.31\\
22 & 203.7 & 1.10 &  99 & 0.29\\
23 & 188.7 & 3.86 & 105 & 0.27\\
24 & 201.6 & 7.58 & 111 & 0.27\\
\hline
\end{tabular}}
\subfloat[$\sim250^{\circ}$C]{
\begin{tabular}{ccccc}
\hline No. & $T_a$& $\dot \varepsilon_a$   & $\sigma_y$& $\varepsilon_{p}$\\
\hline
25 & 249.8 & 0.08 &  95 & 0.42\\
26 & 251.6 & 0.08 &  91 & 0.20\\
27 & 247.5 & 0.08 &  89 & 0.33\\
28 & 257.9 & 0.70 & 101 & 0.44\\
29 & 259.9 & 4.39 & 106 & 0.40\\
30 & 258.5 & 8.24 & 111 & 0.30\\
\hline
\end{tabular}}\\
\subfloat[$\sim300^{\circ}$C]{
\begin{tabular}{ccccc}
\hline No. & $T_a$& $\dot \varepsilon_a$   & $\sigma_y$& $\varepsilon_{p}$\\
\hline
31 & 297.8 & 0.08 &  90 & 0.33\\
32 & 300.5 & 0.09 &  85 & 0.21\\
33 & 307.6 & 0.80 & 101 & 0.47\\
34 & 310.2 & 3.59 & 108 & 0.34\\
\hline
\end{tabular}}
\label{table:Results1}
\end{table}

\begin{table}[!h]
\ContinuedFloat
 \caption{$T_a$, $\dot \varepsilon_a$, $\sigma_y$ and $\varepsilon_p$ attained for all tests (cont'd)}
\centering
\footnotesize
\captionsetup[subfigure]{font=scriptsize,skip=-1pt}
\vspace{-0.125 in}
\subfloat[$\sim350^{\circ}$C]{
\begin{tabular}{ccccc}
\hline No. & $T_a$& $\dot \varepsilon_a$   & $\sigma_y$& $\varepsilon_{p}$\\
\hline
35 & 350.1 & 0.08 &  60 & 0.20\\
36 & 345.3 & 0.09 &  65 & 0.34\\
37 & 355.0 & 0.81 &  71 & 0.47\\
38 & 360.6 & 4.92 &  84 & 0.48\\
39 & 351.9 & 8.82 &  81 & 0.35\\
\hline
\end{tabular}}
\subfloat[$\sim400^{\circ}$C]{
\begin{tabular}{ccccc}
\hline No. & $T_a$& $\dot \varepsilon_a$   & $\sigma_y$& $\varepsilon_{p}$\\
\hline
40 & 400.7 & 0.09 &  43 & 0.22\\
41 & 396.8 & 0.09 &  45 & 0.38\\
42 & 404.3 & 0.83 &  45 & 0.50\\
43 & 407.2 & 4.28 &  56 & 0.48\\
44 & 407.2 & 8.59 &  71 & 0.35\\
45 & 408.7 & 9.17 &  59 & 0.37\\
46 & 404.4 & 9.68 &  61 & 0.15\\
\hline
\end{tabular}}\\
\subfloat[$\sim450^{\circ}$C]{
\begin{tabular}{ccccc}
\hline No. & $T_a$& $\dot \varepsilon_a$   & $\sigma_y$& $\varepsilon_{p}$\\
\hline
47 & 444.4 & 0.10 &  27 & 0.41\\
48 & 452.5 & 0.88 &  33 & 0.52\\
49 & 458.6 & 5.33 &  42 & 0.53\\
50 & 445.5 & 10.35 &  50 & 0.40\\
\hline
\end{tabular}}
\subfloat[$\sim500^{\circ}$C]{
\begin{tabular}{ccccc}
\hline No. & $T_a$& $\dot \varepsilon_a$   & $\sigma_y$& $\varepsilon_{p}$\\
\hline
51 & 494.9 & 0.10 &  17 & 0.40\\
52 & 502.3 & 0.98 &  24 & 0.54\\
53 & 502.6 & 4.97 &  32 & 0.55\\
54 & 498.8 & 11.21 &  39 & 0.46\\
55 & 502.8 & 12.24 &  38 & 0.19\\
\hline
\end{tabular}}
 \label{table:Results2}
\end{table}
\addtolength{\belowcaptionskip}{0pt}

\clearpage
%\begin{table}
%\begin{tabular}{llllllllll}
%\hline
% & \multicolumn{9}{l}{Temperature}\\
% & $30^{\circ}$C & $100^{\circ}$C & $200^{\circ}$C & $250^{\circ}$C & $300^{\circ}$C & $350^{\circ}$C & $400^{\circ}$C & $450^{\circ}$C & $500^{\circ}$C\\
%$\dot \varepsilon \in R_1=$ 0.06 to 0.11/s & 1 & 1 & 1 & 1 & 1 & 1 & 1 & 1 & 1\\
%$\dot \varepsilon \in R_2=$ 0.57 to 0.98/s & 1 & 1 & 1 & 1 & 1 & 1 & 1 & 1 & 1\\
%$\dot \varepsilon \in R_3=$ 3.59 to 5.33/s & 1 & 1 & 1 & 1 & 1 & 1 & 1 & 1 & 1\\
%$\dot \varepsilon \in R_4=$ 7.58 to 12.24/s & 1 & 1 & 1 & 1 & 1 & 1 & 1 & 1 & 1\\
%\hline
%\end{tabular}
%\end{table}

The yield stress for each test (shown in Fig. 3-6 as circle symbols on each curve) was found with a 0.2\% offset method employing temperature-corrected shear and elasticity modulii \cite{Frost.82}, $\mu$ and $E$, according to:
\begin{gather}
\mu=2.54\times10^4\left(1+\frac{300-T}{2T_{\text{melt}}}\right)\\
E=2\mu\left(1+\nu\right)
\end{gather}
\noindent based on pure aluminum, with $T$ in Kelvin and $\nu=0.33$. A relationship based on pure aluminum is appropriate as the temperature dependence of $E$ for aluminum alloys is relatively insensitive to solute content \cite{MeyersChawla.09}. The sparse nature of the experimental data in the elastic region of the flow stress coupled with the diametral strain measurement precludes using experimental data directly for temperature corrected modulii.

The mechanical response of this as-cast material exhibits three temperature-dependent regimes of behaviour. Below 300$^{\circ}$C, the flow stress of the material shows appreciable strain hardening at all strain rates and little strain rate sensitivity.  Between 300-350$^{\circ}$C, the strain hardening diminshes and the strain rate sensitivity increases owing to dynamic recovery. Above 350$^{\circ}$C, there is little to no strain hardening and continued strain rate sensitivity. The strain rate dependence is also demonstrated by the trend in $\sigma_y$ (Table \ref{table:Results2}). For the low temperature tests, $\sigma_y$ remains relatively insensitive to strain rate as compared to tests at elevated temperatures.  For the tests conducted at $\sim30$ and $\sim100^{\circ}$C, $\sigma_y$ for $\dot \varepsilon_a \in R_1,R_2$ was found to be approximately 108 MPa varying by $\pm$10\%. The estimated experimental error in the load transducer ($\pm$0.01kN) used in this study is two orders of magnitude lower than the variance observed in the measured yield stress, thus the majority of the variability in the measured yield stress is attributed to small variations in the SDAS and the coarse Mg$_2$Si precipitates in the as-cast structure. Owing to the low area percent and the fact that the applied load will tend to close pores, porosity is not considered to influence the compressive strength of the material \cite{Estey.04}.

For comparison, the low temperature $\sigma_y$ observed in this study of as-cast material is half that of similar material in the T6 condition \cite{Kearney.90} and twice that of the solutionized condition \cite{Estey.04}. The main cause of this difference is surmised to be the size and distribution of the Mg$_2$Si precipitates. In the solutionized condition, the precipitates have dissolved forming a supersaturated solid solution. This results in an ineffective medium for dislocation storage and a substantively smaller $\sigma_y$ compared to the as-cast condition. The artificial ageing step in the T6 process serves to form smaller, more evenly distributed Mg$_2$Si precipitates as compared to the coarse variety in the as-cast condition. This in turn serves as a more effective dislocation storage mechanism, resulting in a higher $\sigma_y$.

For tests conducted at the lowest strain rate, $R_1$, there is evidence of slight strain-softening at temperatures near 300$^{\circ}$C (Fig. \ref{fig4}), which diminishes at higher temperatures for tests in the same strain rate range. This behavior was also observed in the 300$^{\circ}$C test for $R_2$ (Fig. \ref{fig5}). These results are evidence of the dynamic phenomena known to occur for this alloy system at these temperatures \cite{Puchi.01} such as recovery and strain ageing. As 300$^{\circ}$C has been identified as the start temperature for the transition from strain-dependence, the material behavior may be conservatively characterized such that below 350$^{\circ}$C, the material is dislocation interaction dominated. Above this temperature, it is strain rate/diffusion dominated. As there is a temperature/strain-rate range where both dislocation interaction and diffusion processes are active, a constitutive relationship that traverses this regime will inherently have difficulty accurately describing the flow stress \cite{Liang.99,Puchi.01}.
\section{Constitutive equation development}
While the experimental results may be divided into two discernable regimes, from an analysis standpoint it is necessary to construct an equation or series of equations that successfully predicts the flow stress across both regimes. As outlined in Section \ref{sec:ConstitutiveExpr}, physically-based and phenomenological expressions will be applied to the experimental data to identify the most accurate. The accuracy of applying a physically-based expression to only the diffusion regime will also be presented. The analysis in the proceeding was accomplished through linear least squares fitting where possible and a Nelder-Mead technique when non-linear fitting was necessary. The fitting procedure for each approach will be presented initially, after which the results of each model will be discussed.
\subsection{Kocks-Mecking}
\begin{figure}[h!]
\centering
      \includegraphics[width=4 in]{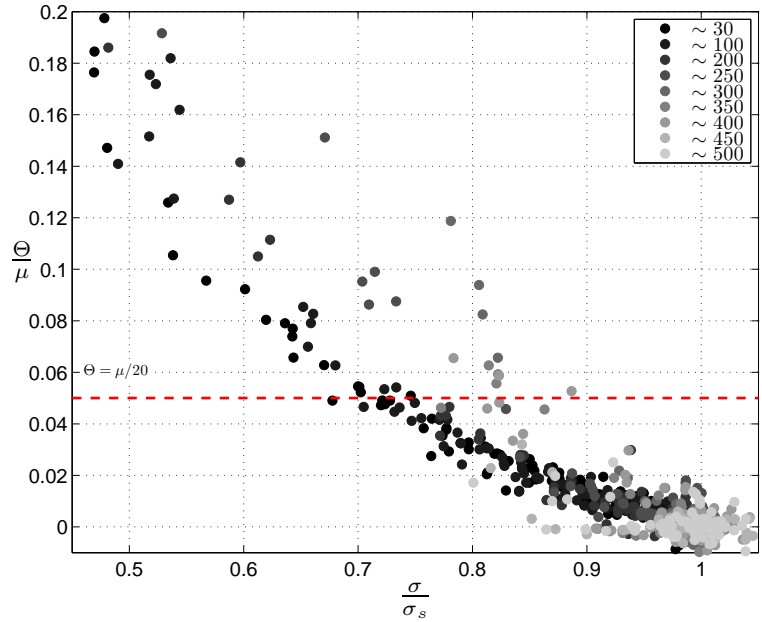}
      \caption[]{$\Theta/\mu$ versus $\sigma/\sigma_s$ for $\dot \varepsilon_a \in R_1$ and $R_4$ at all temperatures. The horizontal line indicates $\Theta=\mu/20$.}
      \label{fig7}
\end{figure}

As outlined in Section \ref{sec:KM_overview}, the Kocks-Mecking description of constitutive behavior consists of relating a saturation stress $\sigma_s$ to a normalized energy term, $g$. Values of $\sigma_s$ were first identified from linear intercepts of $\Theta-\sigma$ data between $\Theta=0$ and $\Theta=\mu/20$, which is defined as the transition from stage III to stage IV \cite{Kuhlmann-Wilsdorf.89,Kocks.03}. The result of this process is demonstrated in Fig. \ref{fig7} which shows $\Theta/\mu$ versus $\sigma/\sigma_s$ for strain rate ranges $R_1$ and $R_4$. The $\Theta/\mu$ versus $\sigma/\sigma_s$ demonstrates overall a continuous function satisfying Eq. \ref{eq:KMbaseline}. While the large strain, high temperature data below $\Theta=\mu/20$ may be approximated by Eq. \ref{eq:KMlinear}, the overall relationship is non-linear. Above this point, the data corresponding to small strains and low temperature represents a large initial work hardening rate. The initial work hardening rate $\Theta_0$ across all tests has been identified as 2875 MPa, as compared to annealed pure aluminum having  $\Theta_0=$ 1120-1720 MPa \cite{Chu.95}.

\begin{figure}[h!]
\centering
      \includegraphics[width=4 in]{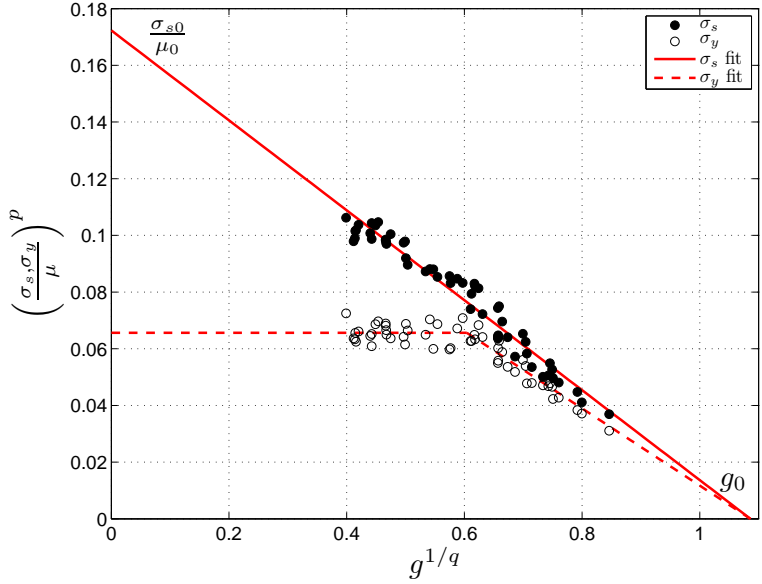}
      \caption[]{Saturation and yield stresses versus normalized activation energy $g$, corrected for obstacle profiles by $p=1/2$ and $q=2$.}
      \label{fig8}
\end{figure}

Fig. \ref{fig8} shows a plot of $\sigma_s/\mu$ versus $g(\dot \varepsilon_a, T_a)$ with $p=1/2$, $q=2$ and $\dot \varepsilon_0=10^7$ s$^{-1}$. These constants are baseline values identified by Kocks and Mecking \cite{Kocks.03}. In Fig \ref{fig7}, the constants $\sigma_{s0}/\mu_0$ and $g_0$ may be determined from the $y$ and $x$ intercepts, respectively. A sensitivity analysis conducted on the values of $p$, $q$ and $\dot \varepsilon_0$ did not show appreciable improvement of the fit within the range defined by their physical basis.

Fig. \ref{fig8} also shows the relationship between $\sigma_y$ and $g$. Coinciding with the two regimes identified in the previous section, there is an identifiable threshold value of $g$ where $\sigma_y$ decreases; $\sigma_y/\mu$ remains constant until this threshold is reached. For larger values of $g$ beyond this transition, $\sigma_y$ is taken to approach $g_0$. In order to characterize this transition, a bi-linear function was fit corresponding to data divided at $g^{1/q}=0.6$:
\begin{equation}\label{eq:KM_sigY}
\sigma_y(g) = \\
\left\{
\begin{array}{lr}
\mu c_y^{1/p}
 & g < \left(\frac{c_y}{C_y}+g_0^{1/q}\right)^{1/p}\\
 \mu\left(C_y\left(g^{1/q}-g_0^{1/q}\right)\right)^{1/p}
& g \geq \left(\frac{c_y}{C_y}+g_0^{1/q}\right)^{1/p} \\
\end{array}
\right.
\end{equation}
where $C_y$ and $c_y$ are constants. The values of all constants are given in Table \ref{table:coefficients}.

\begin{table}[h!]
\centering
\caption{Values of fitted coefficients in each material model\label{table:coefficients}}
%    \begin{singlespacing}
    \begin{tabular}{p{0.75 in}lll}
    \hline
    Model & Eq. & \multicolumn{2}{c}{Coefficients} \\
    \hline
    \multirow{5}{0.75 in}{Kocks-Mecking}   &\multirow{1}{*}{\ref{eq:KM_sigS}} & $\sigma_{s0}/\mu_0=0.0297$ &$g_0=1.1798$\\
                                        \cmidrule(l){2-4}
                                        &\multirow{1}{*}{\ref{eq:KM_sigY}} & $c_y=0.0656$& $C_y=-0.1361$\\
                                        \cmidrule(l){2-4}
                                        &\ref{eq:KMfinal}                  & $\Theta_0=2850$ MPa &\\
    \hline
    \multirow{5}{0.75 in}{Modified Johnson-Cook}& \ref{eq:stdJC} & $\dot \varepsilon^{\ast}=0.7469\text{s}^{-1}$ &\\
                                        \cmidrule(l){2-4}
                                                & \ref{eq:homoTemp} & $T^{\ast}=34.3^{\circ}$C &\\
                                        \cmidrule(l){2-4}
                                                & \ref{eq:JCrefstress} & $\sigma_s^{\ast}=265.7$ MPa& $\sigma_y^{\ast}=108.1$  MPa\\
                                                &                      & $\Theta_0=4024$ MPa&\\
                                        \cmidrule(l){2-4}
                                        &\multirow{2}{*}{\ref{eq:Kt_jc}}& $\kappa_1=1.4125$& $\lambda_1=1.4584$\\
                                        &                           & $\kappa_2=0.4143$ & $\lambda_2=4.5004$\\
                                        \cmidrule(l){2-4}
                                        &\multirow{2}{*}{\ref{eq:JC_cfactor}}& $\theta_{1}=0.312$ &$\theta_{2}=0.023$\\
                                        &                           & $\Lambda=4.0072$& \\
    \hline
    \multirow{6}{0.75 in}{Extended Ludwik-Hollomon}& \multirow{2}{*}{\ref{eq:LH_N}} & $T_t=346.3^{\circ}$C&$a=-5.8165\times10^{-4}$\\
                                        &                           &$b=0.1962$&$c=-5.2734\times10^{-3}$\\
                                        \cmidrule(l){2-4}
                                        & \multirow{2}{*}{\ref{eq:LH_M}} & $\alpha=4.6489\times10^{-11}$& $\beta=3.5237$\\
                                        &                                & $\gamma=0.2371$\\
                                        \cmidrule(l){2-4}
                                        &\multirow{2}{*}{\ref{eq:LH_K}} & $k_1=8.1200\times10^{-4}$& $k_2=-1.1570$\\
                                        &                               & $k_3=407.7$ &\\
 \hline
    \multirow{3}{0.75 in}{Zener-Hollomon}& \multirow{2}{*}{\ref{eq:ZenerSig}} & $\Phi=2.7209\times10^{19}$ & $\rho=0.3700$\\
                                        &                               & $\eta=1.0265$&\\
                                        \cmidrule(l){2-4}
                                        &  \multirow{1}{*}{\ref{eq:ZenerEps}} & $Q=2.3690\times10^{5}$ J&\\
 \hline
\end{tabular}
\end{table}

As both $\sigma_s$ and $\sigma_y$ are able to be scaled as a function of temperature and strain rate, an expression for flow stress is possible through Eq. \ref{eq:KMbaselineflow}. However, this relationship is incapable of describing a large portion of the flow stress accurately, owing to the particularly high initial work hardening rate, and the non-linear strain-hardening rate at low temperatures and strains (Fig. \ref{fig7}). Using a hyperbolic equation to describe the $\Theta-\sigma$ behavior \cite{Estrin.84} has been found to provide a better phenomenological representation of  $\sigma$ versus $\varepsilon$ post yield for all conditions such that:
\begin{equation}\label{eq:KMfinal}
\hat \sigma_{\text{KM}}=\left(\sigma_s^2+\left(\sigma_y^2-\sigma_s^2\right)\exp\left(\frac{-\Theta_0\varepsilon}{\sigma_s}\right)\right)^{1/2}
\end{equation}
Both in this expression and in Eq. \ref{eq:KMbaselineflow}, $\Theta_0$ is taken to be a static value, where $\sigma_{s}$ and $\sigma_{y}$ are based on preselected dislocation parameters ($p$, $q$) for an absolute strain rate of $\dot \varepsilon_0$. This differs from other approaches taken to assessing these parameters for Al-Mg alloys, which involve regressively solving an objective function \cite{Puchi.01}, which can potentially predict negative flow stresses under conditions resulting in low flow stresses. The other caveat of using a static value of $\Theta_0$ is that the slight strain hardening followed by softening discernable at 300$^{\circ}$C is not captured. However, this phenomena is only pronounced for a small range of thermomechanical states. Overall, this approach provides a reasonable description of the entire range of behavior observed without having to scale $\Theta_0$.
\subsection{Johnson-Cook}\label{sec:JC_applied}
As described in Section \ref{sec:JC_overview}, the Johnson-Cook constitutive behavior is based on scaling a reference flow stress expression to account for changes in strain rates and temperatures. In order to identify a reference flow stress expression, the test data was aggregated for samples tested at $\sim30^{\circ}$C ($T^{\ast}= 34.3^{\circ}$C in Eq. \ref{eq:homoTemp}) and $\dot \varepsilon \in R_1$. Owing to the significant initial work hardening rate as described in the previous section, a single value of $n^{\ast}$ does not adequately describe the work hardening rate at all strains. Therefore, a hyperbolic Voce-type saturation stress model of the same form as Eq. \ref{eq:KMfinal} was substituted for $\sigma^{\ast}$ (Eq. \ref{eq:stdJC}):
\begin{equation}\label{eq:JCrefstress}
\sigma^{\ast}(\varepsilon)=\left((\sigma_s^{\ast})^2+\left((\sigma_y^{\ast})^2-(\sigma_s^{\ast})^2\right)\exp\left(\frac{-\Theta_0\varepsilon}{\sigma_s^{\ast}}\right)\right)^{1/2}
\end{equation}
where $\sigma_{s}^{\ast}$ and $\sigma_{y}^{\ast}$ coincide with the average values for $\dot \varepsilon_a \in R_1$ at $\sim30^{\circ}$C. This approach expands on the linear Voce-type characterization that was previously applied by Alkarca et al. \cite{Alkarca.07} in an amended Johnson-Cook type model for A356 in the T6 condition. The reference strain rate was taken to be the overall average strain rate of $R_1$.

The thermal softening function $f(T)$ was derived by rearranging Eq. \ref{eq:stdJC} to give:
\begin{equation}\label{eq:CoeffExtractJC}
\frac{\sigma}{\sigma^{\ast}(\varepsilon)}=f(T)+f(T)C\ln\left(\frac{\dot \varepsilon}{\dot \varepsilon^{\ast}}\right)
\end{equation}
Values for $f(T)$ may be determined as the linear intercepts of $\sigma/\sigma^{\ast}$ versus $\ln(\dot \varepsilon/\dot \varepsilon^{\ast})$. The values of $f(T)$ were calculated from the reference tests for strains between 0.05 to 0.12 at increments of 0.01.  This strain range was selected because it represents the post-yield response and terminates prior to the peak strain $\varepsilon_p$ attained for all tests.

\begin{figure}
  \centering
  \subfloat[Thermal softening behavior]{\label{fig:KTJC}\includegraphics[width=4 in]{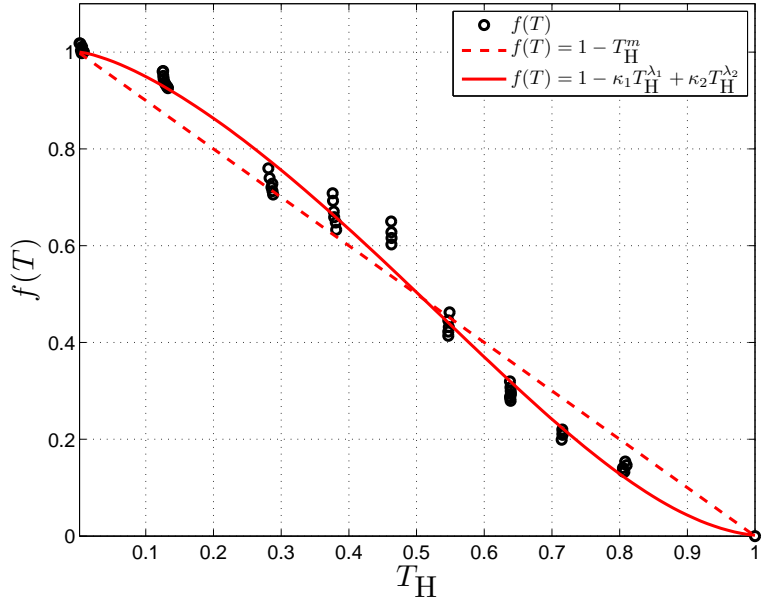}}\\
  \subfloat[$C$ values]{\label{fig:CJC}\includegraphics[width=4 in]{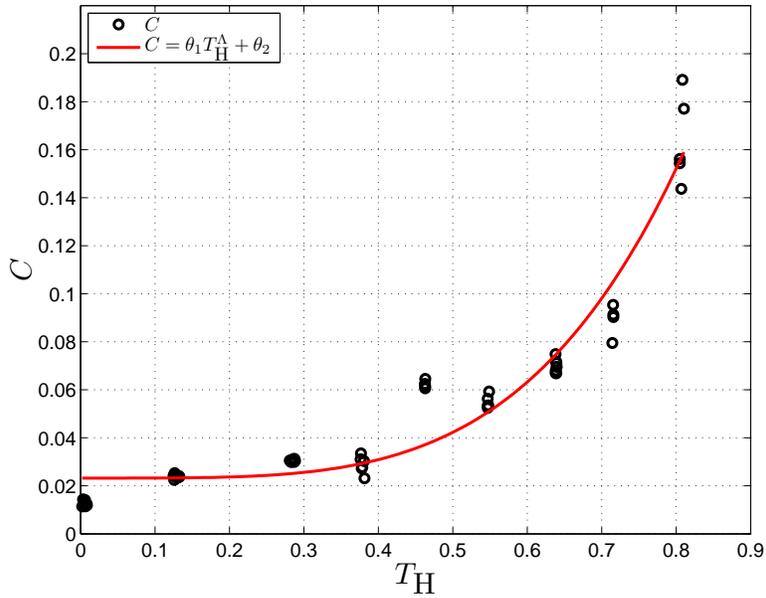}}
  \caption{Johnson-Cook parameters plotted versus homologous temperature.}
  \label{fig:NMK}
\end{figure}

Comparing the standard form of $f(T)=1-T_{\text{H}}^m$ (Eq. \ref{eq:stdJC}) with the calculated values of $f(T)$ versus the homologous temperature $T_H$  indicates that $m$ is greater than 1 at low temperatures where strain hardening is dominant, but is less than 1 at elevated temperatures where strain rate effects are dominant. As a result, a single coefficient of $m$ does not adequately capture the thermal softening term. Therefore, the standard expression for $f(T)$ has been expanded such that there are two terms to describe each of the ranges of behavior:
\begin{equation}\label{eq:Kt_jc}
f(T)=1-\kappa_1T^{\ast \lambda_1}+\kappa_2T^{\ast \lambda_2}
\end{equation}
where $\kappa_1$ and $\lambda_1$ apply for low temperatures while $\kappa_2$ and $\lambda_2$ describe elevated temperatures. The reference temperature $T^{\ast}$ was taken to be the minimum temperature found for all strains. A comparison between Eq. \ref{eq:Kt_jc} and $f(T)=1-T_{\text{H}}^m$ is shown in Fig. \ref{fig:KTJC}.

The strain rate coefficient $C$ was then determined by dividing the slope term in Eq. \ref{eq:CoeffExtractJC} by Eq. \ref{eq:Kt_jc}. Plotting $C$ versus $T_H$ (Fig. \ref{fig:CJC}) shows that this term increases exponentially with temperature past a critical homologous temperature such that:
\begin{equation}\label{eq:JC_cfactor}
C=\theta_1T_{\text{H}}^{\Lambda}+\theta_2\\
\end{equation}
%The modified Johnson-Cook model as applied to as-cast A356 is therefore:
%\begin{multline}
%\hat \sigma_{\text{JC}}=\left((\sigma_s^{\ast})^2+\left((\sigma_y^{\ast})^2-(\sigma_s^{\ast})^2\right)\exp\left(\frac{-\Theta_0\varepsilon}{\sigma_s^{\ast}}\right)\right)^{1/2}\\
%\left(1+\left(\theta_1T_{\text{H}}^{\Lambda}+\theta_2\right)\ln\left(\frac{\dot \varepsilon}{\dot \varepsilon^{\ast}}\right)\right)\\
%\left(1-\kappa_1T_{\text{H}}^{ \lambda_1}+\kappa_2T_{\text{H}}^{ \lambda_2}\right)
%\end{multline}
%where $\sigma_{s,y}^{\ast}$ coincide with the average values for $\dot \varepsilon_a \in R_1$ at $\sim30^{\circ}$C.

\subsection{Extended Ludwik-Hollomon}
\begin{figure}
  \centering
  \subfloat[$N$ and $M$ values]{\label{fig:nmHollo}\includegraphics[width=4 in]{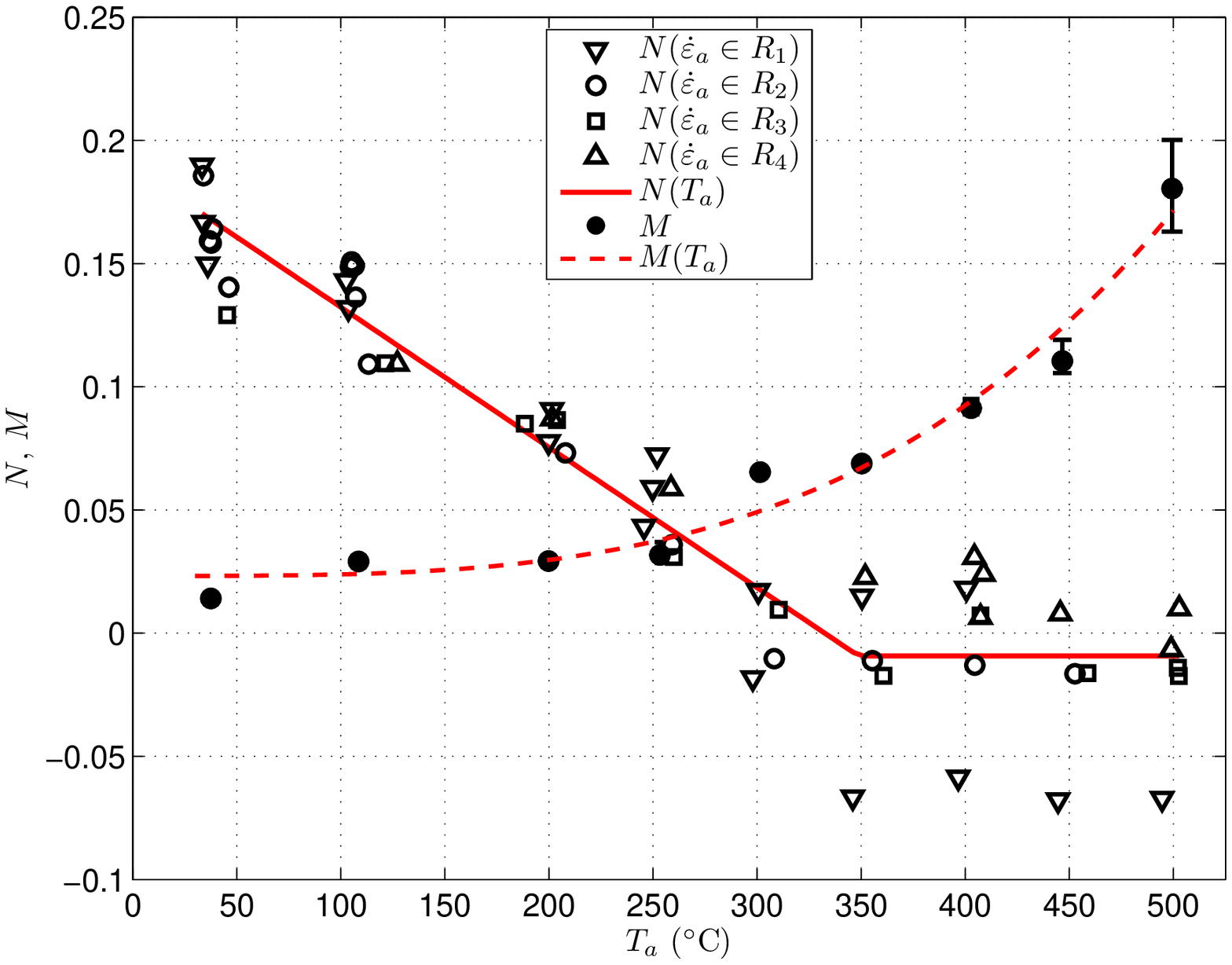}}\\
  \subfloat[$K$ values]{\label{fig:KHollo}\includegraphics[width=4 in]{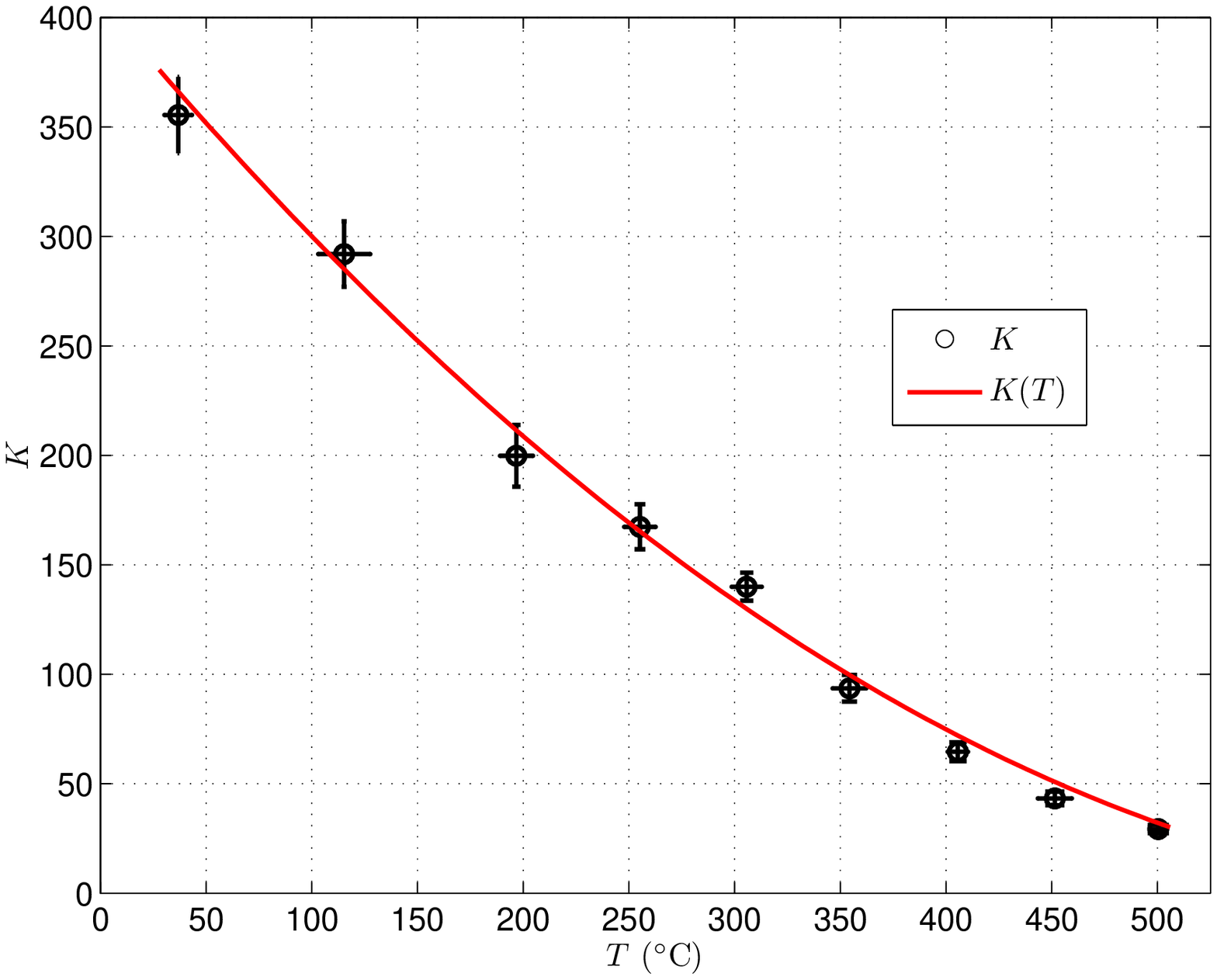}}
  \caption{Extended Hollomon coefficients plotted versus temperature.}
  \label{fig:NMK}
\end{figure}

Outlined in Section \ref{sec:LH_overview} and Eq. \ref{eq:LHbaseline}, the extended Ludwik Hollomon expression for flow stress is coupled with temperature via strength, work hardening and strain rate parameters ($K$,$N$,$M$). Values of the work hardening parameter, $N$, were found first via the slope of $\ln\sigma$-$\ln\varepsilon$ curves. The strain-rate term, $M$, was taken to be the slope of $\ln\sigma$-$\ln\dot\varepsilon$ curves built from data at the same strains used in the Johnson-Cook analysis. The resulting values of $M$ and $N$ versus $T_a$ are shown in Fig. \ref{fig:nmHollo}. These two data sets were then fit with Eq. \ref{eq:LH_N} and \ref{eq:LH_M} and were used to extract values for $K$ (Fig. \ref{fig:KHollo}).

The $N$ values plotted versus average temperature demonstrate that there is a roughly linear trend for all strain rates until approximately $350^{\circ}$C. Here, the diffusion effects identified in Section \ref{Sec:ExpResults} become significant. The values for $\dot \varepsilon_a \in R_1$ are distinctly negative, while the rest of the strain rates result in $N$ values close to zero. Corresponding to regimes exhibiting strain-rate dependent phenomena, $N(T)$ was characterized such that:
\begin{equation}\label{eq:LH_N}
N(T) = \\
\left\{
\begin{array}{lr}
aT+b
 & T\leq T_t\\
 c
& T> T_t \\
\end{array}
\right.
\end{equation}
where $T_t$ is the transition temperature ($\sim350^{\circ}$C) between the two regimes, with $a$, $b$ and $c$ constants fit using values of $T_a\geq300^{\circ}$C and $T_a<300^{\circ}$C respectively.

As the physical basis for the trend in the $M-T_a$ data (Fig. \ref{fig:nmHollo}) is an observed increase in strain rate sensitivity with temperature, a continuous function similar to that used for $C$ (Fig. \ref{fig:CJC}) has been employed for $M(T)$:
\begin{equation}\label{eq:LH_M}
M(T)=\alpha T^{\beta}+\gamma
\end{equation}
The correlation of the strength coefficient with temperature, $K(T)$, is continuous and was found to be best described by:
\begin{equation}\label{eq:LH_K}
K(T)=k_1T^2+k_2T+k_3
\end{equation}
\subsection{Zener-Hollomon}
\begin{figure}[h!]
\centering
      \includegraphics[width=4 in]{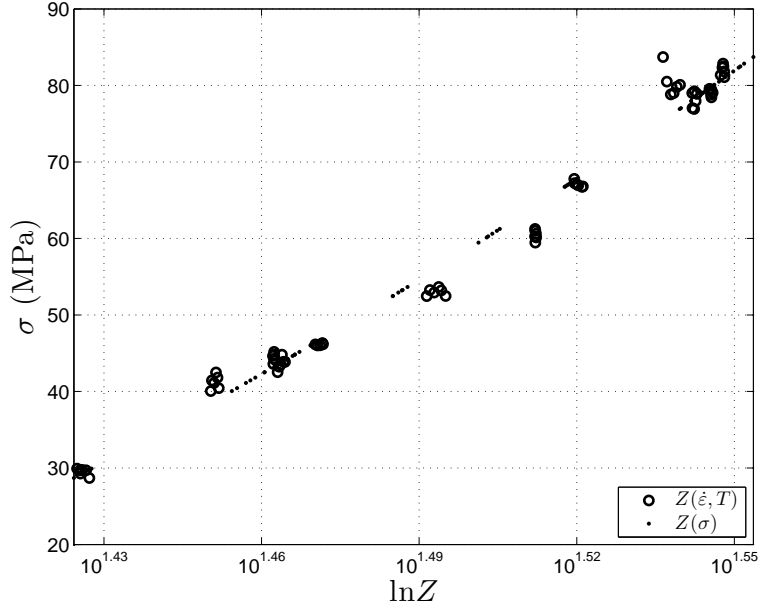}
      \caption[]{$Z=f(\dot \varepsilon, T)$ compared to $Z=f(\sigma)$.}
      \label{fig11}
\end{figure}

For the Zener-Hollomon analysis outlined in Section \ref{sec:ZH_overview}, it was assumed that the diffusion behavior at low strain rates is negligible. The Zener-Hollomon parameters were determined with $T_a\geq350^{\circ}$C. The constants in Eq. \ref{eq:ZenerSig} and \ref{eq:ZenerEps} were fitted by starting with an initial value of $\rho$, which was then used to find $\eta$, followed by $Q$ and $\Phi$. The value of $\rho$ was then iterated upon to converge on a single value of $Q$ at each strain between 0.1 and 0.15 in increments of 0.01. Values of $\eta$ were found as the reciprocal of the slope of $\ln( \sinh(\rho \sigma))$ versus $\ln (\dot \varepsilon)$ \cite{Garafalo.63}. By rearranging Eq. \ref{eq:ZenerEps} and \ref{eq:ZenerSig}:
\begin{equation}
\ln (\sinh (\rho \sigma))-\frac{\ln \dot \varepsilon}{\eta}=\frac{Q}{\eta R T}-\frac{\ln (\Phi)}{\eta}
\end{equation}
Iterating over a range of $\rho$ provides values of $Z$ as a function of strain rate, temperature and stress (Fig. 11), converging at $Q=237$ kJ/mol (Fig. \ref{fig11}). In order to assess the sensitivity of the technique, the data from high-temperature tests ($\geq\sim400^{\circ}$C) for select strain rates ($\dot \varepsilon \in R_{2-4}$) was fit in the same manner. This data set was selected because it does not exhibit softening effects. The activation enthalpy, $Q$, for this data set was found to be 157 kJ/mol, which is considerably closer to the activation energy for self-diffusion of aluminum and diffusion of magnesium in aluminum, 131 and 137 kJ/mol respectively.

\section{Comparison of constitutive expressions and discussion}
\begin{figure}
  \centering
  \subfloat[$\sim100^{\circ}$C]{\label{fig:comp100}\includegraphics[width=0.5\textwidth]{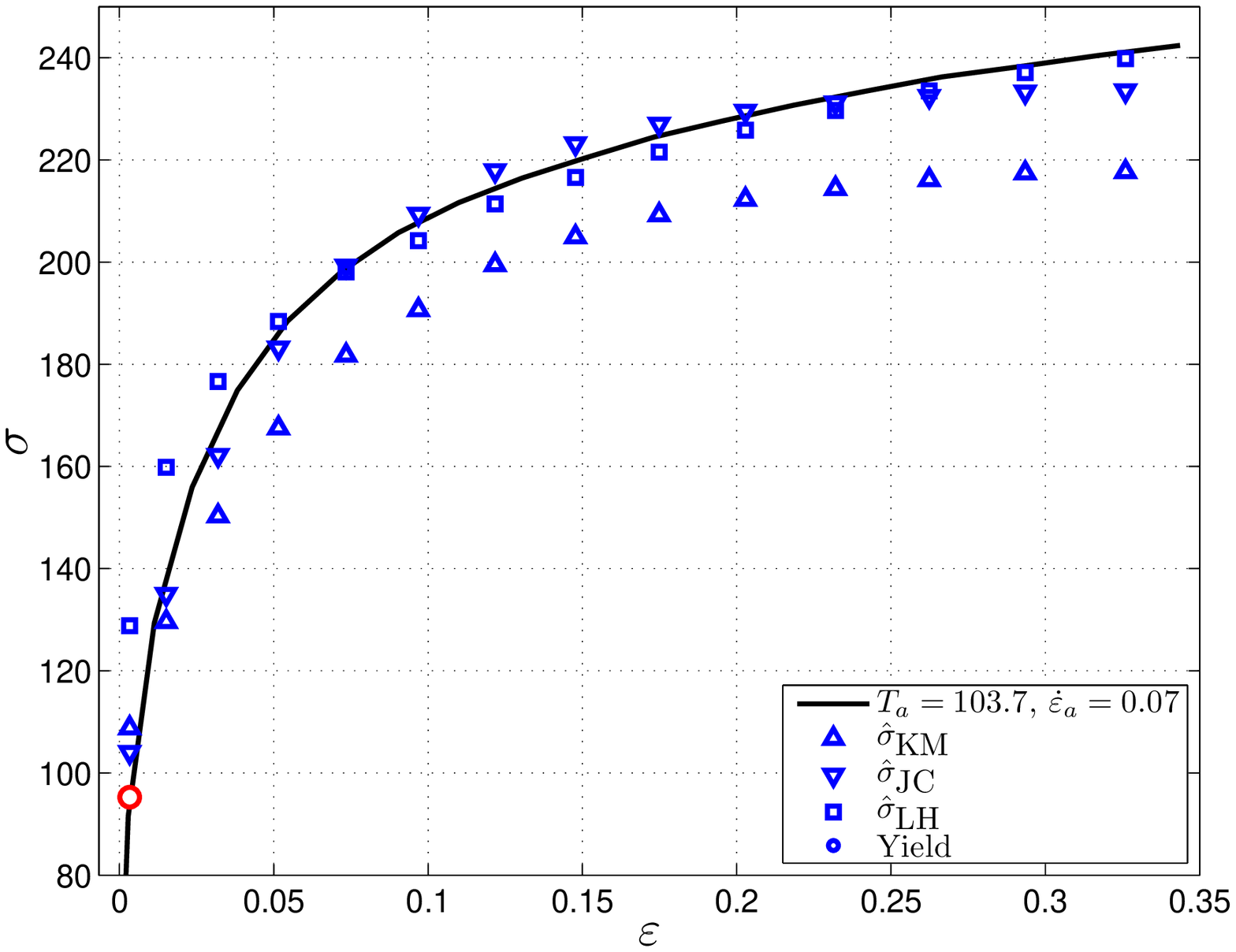}}
  \subfloat[$\sim200^{\circ}$C]{\label{fig:comp200}\includegraphics[width=0.5\textwidth]{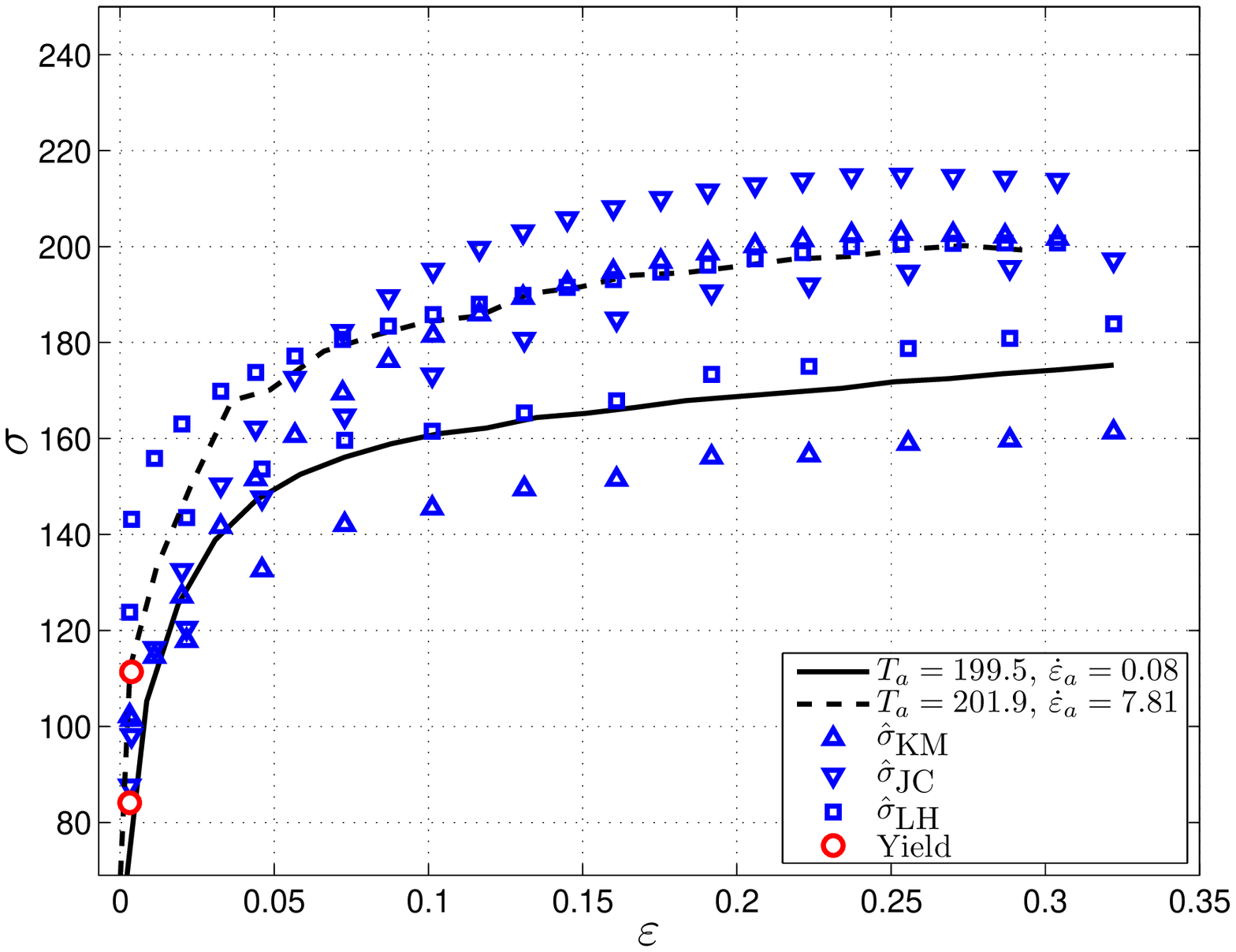}}\\
  \subfloat[$\sim450^{\circ}$C]{\label{fig:comp450}\includegraphics[width=0.5\textwidth]{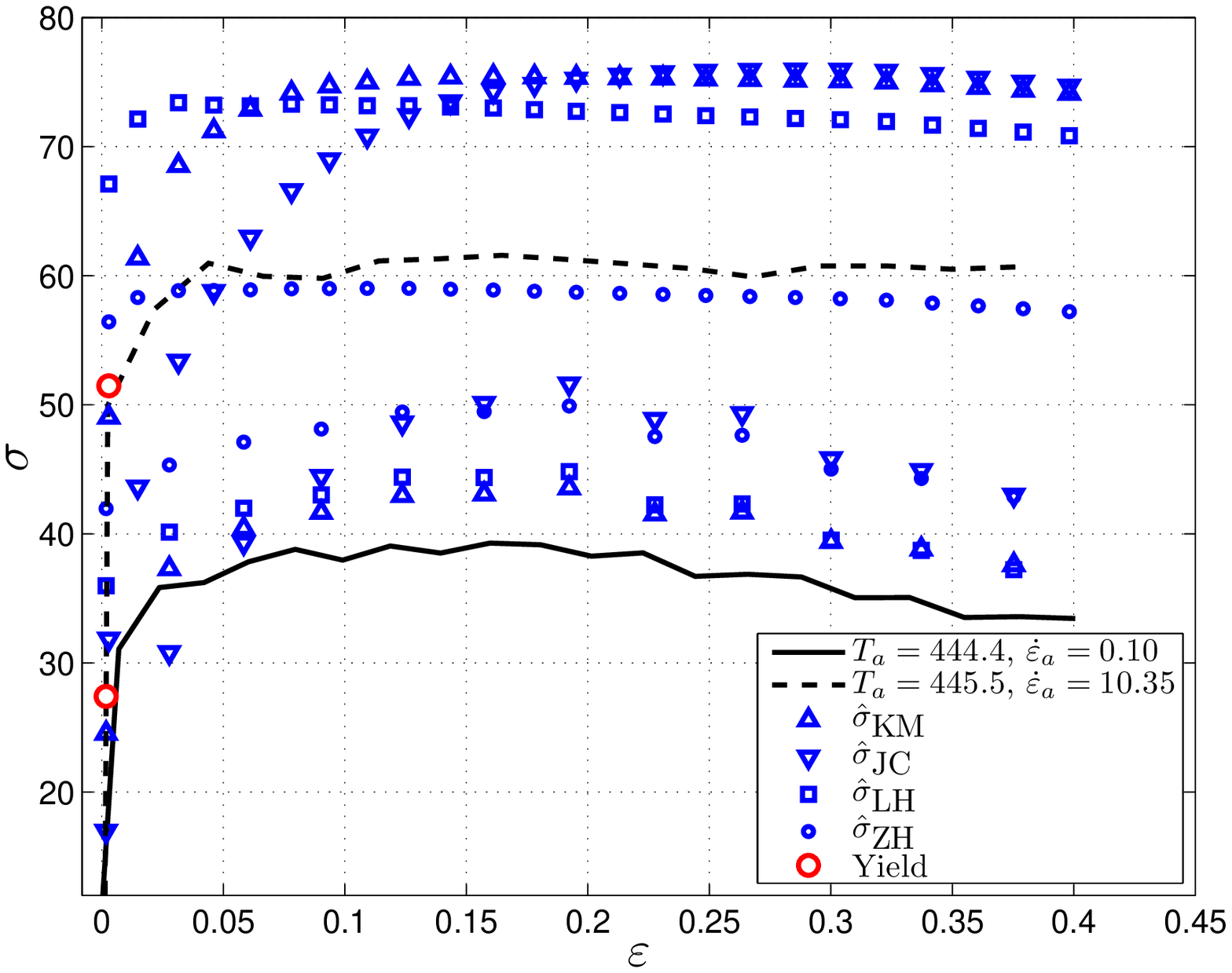}}
  \subfloat[$\sim500^{\circ}$C]{\label{fig:comp500}\includegraphics[width=0.5\textwidth]{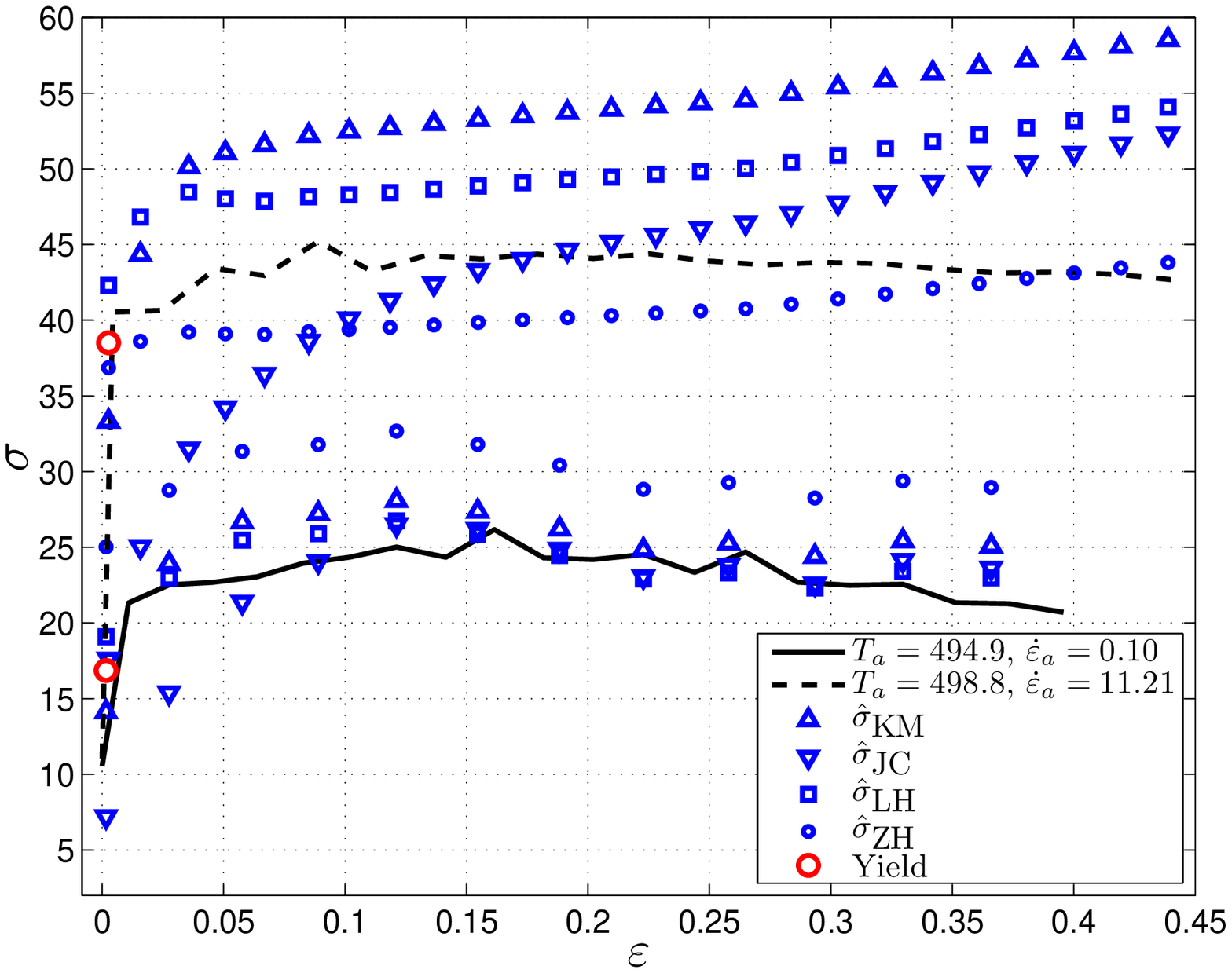}}
  \caption{Comparison of experimental versus predicted flow stress for applicable models.}
  \label{fig:compare}
\end{figure}

Fig. \ref{fig:compare} shows examples of the measured flow stress compared to the predicted flow stress calculated by the different constitutive expressions. In these calculations, experimental data points for strain, strain rate and temperature are directly substituted into each respective constitutive expression for flow stress. All constitutive relationships predict the measured stresses reasonably well considering material strength variations ($\sim10\%$)  Figures \ref{fig:comp100} and \ref{fig:comp200} demonstrates the characterization of the Kocks-Mecking, Johnson-Cook and Ludwik-Hollomon expressions at temperatures that exhibits strain hardening (100$^{\circ}$C and 200$^{\circ}$C), while Figures \ref{fig:comp450} and \ref{fig:comp500} shows these expressions as well as the Zener-Hollomon characterization for the strain rate sensitive regime (450$^{\circ}$C and 500$^{\circ}$C). Owing to the negligable strain rate sensitivity at 100$^{\circ}$C, a single strain rate is presented corresponding to $R_1$ in Fig. 12 \ref{fig:comp100}, while the latter plots in Fig. 12 are representative of the lowest and highest experimental strain rate ranges, $R_1$ and $R_4$.

At lower temperatures, all applicable relationships describe the experimental flow stress well. The Ludwik-Hollomon fits the data best for these conditions, fitting the data for both strain rates particularly well at 200$^{\circ}$C (Fig. \ref{fig:comp200}). The Kocks-Mecking underestimates the low strain rate flow stress at both 100$^{\circ}$C and 200$^{\circ}$C, but is acceptable for the elevated strain rate at 200$^{\circ}$C. The Johnson-Cook expression overestimates the flow stress in all cases, a trend that continues at elevated temperatures.

At elevated temperatures, the Ludwik-Hollomon, Kocks-Mecking and Johnson-Cook expressions are fairly accurate at low strains, but overestimate the flow stress at elevated strain rates. The Zener-Hollomon expression captures the experimental data at all strain rates compared to the other three. Of the three expressions capable of capturing strain hardening, the Johnson-Cook is clearly incorrect even from a phenomenological standpoint: it demonstrates significant hardening even at 500$^{\circ}$C (Fig. \ref{fig:comp500}).

Calculating the sum of the mean difference as well as the mean square error (MSE) between the predicted flow stress using experimental data points and the experimental flow stress measured directly for all tests conducted provides a measure of each models performance. For the constitutive expressions applicable across both regimes, the Kocks-Mecking slightly underestimates the flow stresses by less than 1\% with 9\% MSE, the Johnson-Cook and Ludwik-Hollomon overestimate the flow stress by 8\% and 2\%, respectively. The overall MSE of the Johnson-Cook expression is 10\%, however the MSE of the Ludwik-Hollomon expression is less than the Kocks-Mecking expression at 6\%. The Zener-Hollomon model overestimates the applicable flow stress by 5\% and has a MSE of 5\% for all experimental data where it has been fit. In terms of specific strain rate and temperature ranges best described by each expression, the general trend for the strain-hardening conditions is that higher error is observed with increasing temperature, with peak error occurring in the transition regime (300-350$^{\circ}$C). This is especially true for the Johnson-Cook expression, where the error reaches 25\% for the 300-350$^{\circ}$C range, while the others were approximately equal to their MSE values.

As this material exhibits a range of temperature and strain rate dependent deformation response, selecting a single constitutive expression to capture the full range of behavior observed requires compromises in accuracy. Even though the Johnson-Cook model is widely used to generally characterize constitutive behavior, applied to as-cast A356, it is not an effective phenomenological approach. This is despite the heavy modification applied in this study to improve its performance (Section \ref{sec:JC_applied}). The extended Ludwik-Hollomon model provides additional flexibility in a phenomenological approach, however requires a large number of fitting coefficients to approximate the flow stress effectively. The Zener-Hollomon technique describes the flow stress effectively, but is only applicable in the rate-dependent regime. The physically-based Kocks-Mecking is more efficient in terms of fitting coefficients, however it requires well characterized yield and saturation stresses. Furthermore, the accuracy of the Kocks-Mecking expression at low strains is dependent on the integrated form of Eq. \ref{eq:KMbaseline} in order to revert to consider strain dependence. The Kocks-Mecking expression in the integrated form may only describe the strain-softening behavior of the material by scaling the initial strain hardening rate according to temperature and strain rate. This requires a large number of tests to accurately assess, which may also be a limit when considering inherent material variations. As an alternative to this physically-based model, the Ludwik-Hollomon approach offers a phenomenological description which has a greater degree of freedom to capture specific material behavior.
\section{Summary and Conclusions}
The constitutive behavior of as-cast A356 has been experimentally characterized through an extensive set of compression tests. The data was used to fit four phenomenological and physically-based constitutive expressions. The material in the as-cast form shows a diverse range of thermomechanical behavior characteristic of Al-Si-Mg alloys, transitioning from strain-hardening to being completely strain rate dependent. Below 350$^{\circ}$C, the material was observed to strain harden, whereas above this temperature it becomes more strain-rate dependent with increased temperature as outlined in Section 3. This is also demonstrated in Section 5, most prominently by the work-hardening coefficient, $N$, plotted versus temperature (Fig. 10a) in the extended Ludwik model. Here, it is effectively 0 at 300$^{\circ}$C. Furthermore, the inflection point of $f(T)$ in the Johnson-Cook model (Fig. 9a) corresponds to 300-350$^{\circ}$C. Finally, $g^{1/q}=0.06$ in Fig. 8 also corresponds to this range of 300-350$^{\circ}$C, where $\sigma_y$ approaches $\sigma_s$.

This behavior poses a challenge for developing a single constitutive expression that accurately describes all experimental results across the temperature and strain rate ranges encompassed experimentally. As a result, each constitutive expression has differing degrees of accuracy and efficiency in predicting flow stress for particular ranges of temperature, strain rate and strain. Specifically, over the experimental data tested:
\begin{itemize}
\item The extended Ludwik-Hollomon expression is the most flexible and therefore provides the best prediction across all temperatures and strain-rates. This expression overestimated the flow stress by an average 2\% with a MSE of 6\%. This phenomenological approach necessitated a large number of fitting coefficients.
\item The Kocks-Mecking relationship on average slightly underestimated the flow stress by 1\%, but with a larger MSE. Flow stresses for elevated strain and strain rate conditions showed better agreement with the relationship. This model is physically-based, and does not have as many fitted coefficients as the Ludwik-Hollomon.
\item The Johnson-Cook expression, did not describe the experimental flow stress effectively, even with heavy modification. Not only did the relationship overpredict the flow stress, it exhibited strain-hardening at all temperatures.
\item For temperatures greater than 400$^{\circ}$C where the material is completely strain-rate dependent, the Zener-Hollomon relationship shows a much better fit to experimental data at all strain rates as compared to the other models. However, the Zener-Hollomon expression is not valid at lower temperatures where strain hardening occurs.
\end{itemize}
Based on this analysis, it has been found that the constitutive expression that most accurately describes as-cast A356 across all temperatures and strain rates is the extended Ludwik-Hollomon expression.
\end{document}

%% file: 12_02_03_Nomenclature.tex
\section*{Nomenclature}
\begin{description}
\item[$\alpha$] strain rate function coefficient
\item[$\beta$] strain rate function coefficient
\item[$\gamma$] strain rate function coefficient
\item[$\varepsilon$] true strain
\item[$\varepsilon_p$] total plastic strain
\item[$\dot \varepsilon$] strain rate (s$^{-1}$)
\item[$\dot \varepsilon_0$] absolute strain rate (s$^{-1}$)
\item[$\dot \varepsilon_a$] average strain rate (s$^{-1}$)
\item[$\dot \varepsilon_{\ast}$] Johnson-Cook reference strain rate (s$^{-1}$)
\item[$\eta$] Zener-Hollomon fit coefficient
\item[$\theta_{1-2}$] Johnson-Cook strain rate adjustment coefficients
\item[$\Theta$] strain hardening rate (MPa)
\item[$\Theta_0$] initial strain hardening rate (MPa)
\item[$\kappa_{1-2}$] Johnson-Cook thermal softening adjustment coefficients
\item[$\lambda_{1-2}$] Johnson-Cook thermal softening adjustment coefficients
\item[$\Lambda$] Johnson-Cook strain rate adjustment coefficient
\item[$\mu$] temperature corrected shear modulus (MPa)
\item[$\mu_0$] absolute shear modulus (MPa)
\item[$\rho$] Zener-Hollomon fit coefficient
\item[$\sigma$] recorded true stress (MPa)
\item[$\hat \sigma$] predicted flow stress (MPa)
\item[$\sigma_s$] saturation or Voce stress (MPa)
\item[$\sigma_{s0}$] absolute saturation stress (MPa)
\item[$\sigma_{\ast}$] Johnson-Cook reference stress (MPa)
\item[$\sigma_y$] yield stress (MPa)
\item[$\Phi$] Zener-Hollomon fit coefficient
\item[$a$,$b$,$c$] extended Ludwik-Hollomon strain hardening function coefficients
\item[$\bar{b}$] burgers vector magnitude (m)
\item[$C$] Johnson-Cook strain rate coefficient
\item[$C_y$,$c_y$] yield strength transition coefficients
\item[$E$] temperature corrected modulus of elasticity (MPa)
\item[$g$] normalized activation energy
\item[$g_0$] initial normalized activation energy
\item[$k$] Boltzmann constant ($1.3806503 \times 10^{-23}$ $\text{m}^2$ $\text{kg}$ $\text{s}^{-2}$ $\text{K}^{-1}$)
\item[$k_{1-3}$] extended Ludwik-Hollomon strength parameter function coefficients
\item[$K_T$] Johnson-Cook thermal softening coefficient
\item[$K(T)$] extended Ludwik-Hollomon strength parameter function
\item[$M(T)$] extended Ludwik-Hollomon strain rate function
\item[$N(T)$] extended Ludwik-Hollomon strain hardening function
\item[$p$] dislocation interaction characteristic parameter
\item[$Q$] Zener-Hollomon activation energy (J)
\item[$q$] dislocation interaction characteristic parameter
\item[$R$] gas constant (8.3144 J/mol$\cdot$ K)
\item[$T$] instantaneous temperature ($^{\circ}$C)
\item[$T^{\ast}$] Johnson-Cook reference temperature ($^{\circ}$C)
\item[$T_a$] average temperature ($^{\circ}$C)
\item[$T_{\text{H}}$] homologous temperature
\item[$T_{\text{melt}}$] melting temperature ($^{\circ}$C)
\item[$T_t$] strain rate coefficient transition temperature ($^{\circ}$C)
\item[$Z$] Zener-Hollomon parameter
\end{description} 